\documentclass [12pt]{article}
\usepackage{graphics,color,epsfig}

\begin{document}

\begin{flushright}
CLNS 00/1706
\end{flushright}
\bigskip
\begin{center}
\textbf{\Large Novel Method of Measuring Electron Positron Colliding Beam
Parameters}\\

\bigskip

D. Cinabro, K. Korbiak\\
\smallskip
{\it Department of Physics and Astronomy,
Wayne State Univeristy,\\
Detroit, MI 48202, USA}\\
\medskip
R. Ehrlich, S. Henderson, N. Mistry\\
\smallskip
{\it Laboratory of Nuclear Studies, Cornell Univeristy,\\
Ithaca, NY 14853, USA}\\

\bigskip

(29 November 2000)
\end{center}

\begin{abstract}
Through the simultaneous measurement of the transverse size 
as a function of longitudinal position, and the 
longitudinal distribution of luminosity, we are able to
measure the $\beta_y^\ast$ (vertical envelope function at
the collision point),
vertical emittance, and bunch length of colliding
beams at the Cornell Electron-positron Storage Ring (CESR).  This 
measurement is possible due to the significant ``hourglass'' effect at 
CESR and the excellent tracking resolution of the CLEO detector.
\end{abstract}
\begin{flushleft}
\end{flushleft}

\bigskip

\begin{center}
{\small
Submitted to Nuclear Instruments and Methods in Physics
Research Section A:  Accelerators, Spectrometers, Detectors and Associated
Equipment}
\end{center}

\newpage

	One of the difficult problems in colliding beam physics is the 
measurement of beam parameters at the collision point.  In this letter, 
we present a method of making such a measurement using precision 
measurements of the luminous region with $e^+e^- \rightarrow \mu^+\mu^-$ 
events detected by a general purpose high energy physics experiment.  
Key to this measurement is the detailed geometry of the highly focused
colliding beams, which leads to the ``hourglass''
effect.\cite{hourglass}

	Tightly focused beams have a ``waist'' at the focal point of the final 
quadrupoles and their size grows away from this waist.  The transverse 
beam size is given by
\begin{equation}
\sigma(z) = \sqrt{\epsilon\beta(z)}
\end{equation}
where $\beta(z)$ is the amplitude or beta function, which depends
on the longitudinal position, $z$, of the beam and 
the emittance, $\epsilon$, which is independent of $z$.  Near a waist, 
$\beta(z)$ can be written as
\begin{equation}
\beta(z) = \beta^\ast + \frac{(z - z_{0 {\rm beta}})^2}{\beta^\ast}
\end{equation}
where $\beta^\ast$ is the value of the beta function at the waist and
$z_{0 {\rm beta}}$  is the longitudinal position of the waist.  Thus the
beam in the
longitudinal and  transverse dimensions forms an hourglass shape with a minimum
size at $z_{0 {\rm beta}}$.

	The hourglass effect arises from the beam being in Gaussian 
shaped bunches with length $\sigma_z$.  If the bunch length is long 
compared with the dimension of the waist, then little of the beam is 
colliding where the beam is narrowest.  In this case,
the longitudinal distribution of luminosity depends not only on
$\sigma_z$, but also on the horizontal and vertical value of the
beta function at the interaction point,
$\beta^\ast_x$ and $\beta^\ast_y$, respectively.  In addition if either
$\beta^\ast$ is smaller than $\sigma_z$, luminosity does not
improve as much as naively expected by making $\beta^\ast$ smaller.
 
	The luminous region is defined by the overlap integral of two 
beams.  Thus we expect the vertical width of the luminous region 
as a function of the longitudinal position to be given by
\begin{equation}
\sigma_y(z) = \sqrt{\frac{\epsilon_y}{2} \left(\beta_y^\ast +
                    \frac{(z - z_{0 {\rm beta}})^2}{\beta_y^\ast}\right)}\ ,
\label{eq:houryres}
\end{equation}
and similarly for the horizontal width.  It is assumed that 
the emittances and $\beta^\ast$'s are the same for the two beams.  The 
beam parameters for the Cornell Electron-positron Storage Ring (CESR) 
for the data discussed in this paper
are given in Table~\ref{tab:beam}.
\begin{table}
\caption{CESR beam parameters at zero bunch current
during the time this measurement was made.}
\begin{center}
\begin{tabular}{|c|c|}
\hline
Parameter    & Value ($\mu$m) \\ \hline
$\beta_x^\ast$    & $1.1996 \times 10^6$ \\
$\epsilon_x$ & 0.21 \\
$\beta_y^\ast$    & 17900  \\
$\epsilon_y$ & 0.0010 \\
$\sigma_z$   & 18100  \\  \hline
\end{tabular}
\end{center}
\label{tab:beam}
\end{table}
All the parameters in Table~\ref{tab:beam} 
are given at zero bunch current.  They are all expected to depend on the
bunch current.  
We have previously observed that $\beta_x^\ast$ 
is reduced by roughly a factor of two in colliding beam conditions
due to the dynamic beta effect.\cite{dybeta}
Likewise, $\beta_y^\ast$ is expected to be
reduced by about 25\% due to beam-beam focusing.  The vertical emittance
depends on the beam-beam tuneshift parameter; for operation in the
saturated tuneshift regime, the vertical beam size increases linearly
with bunch current.  Additionally, there are streak camera observations
which show an increase in $\sigma_z$  as the bunch current
increases.\cite{bunchl} 
The method described in this paper is aimed at measuring some of
these dynamic effects by direct observation of the luminous region.

	Note that these measurements are taken over a long time,
about four months of CESR and CLEO running, and at many different
machine conditions.  Thus we expect only rough agreement with the
parameters given in Table~\ref{tab:beam}, but we should be
sensitive to the dynamic effects discussed above.  We expect
to observe a larger $\epsilon_y$ and $\sigma_z$, and a smaller
$\beta_y^\ast$ and $\beta_x^\ast$ than given in Table~\ref{tab:beam}.

\begin{figure}
\begin{center}
\epsfig{file=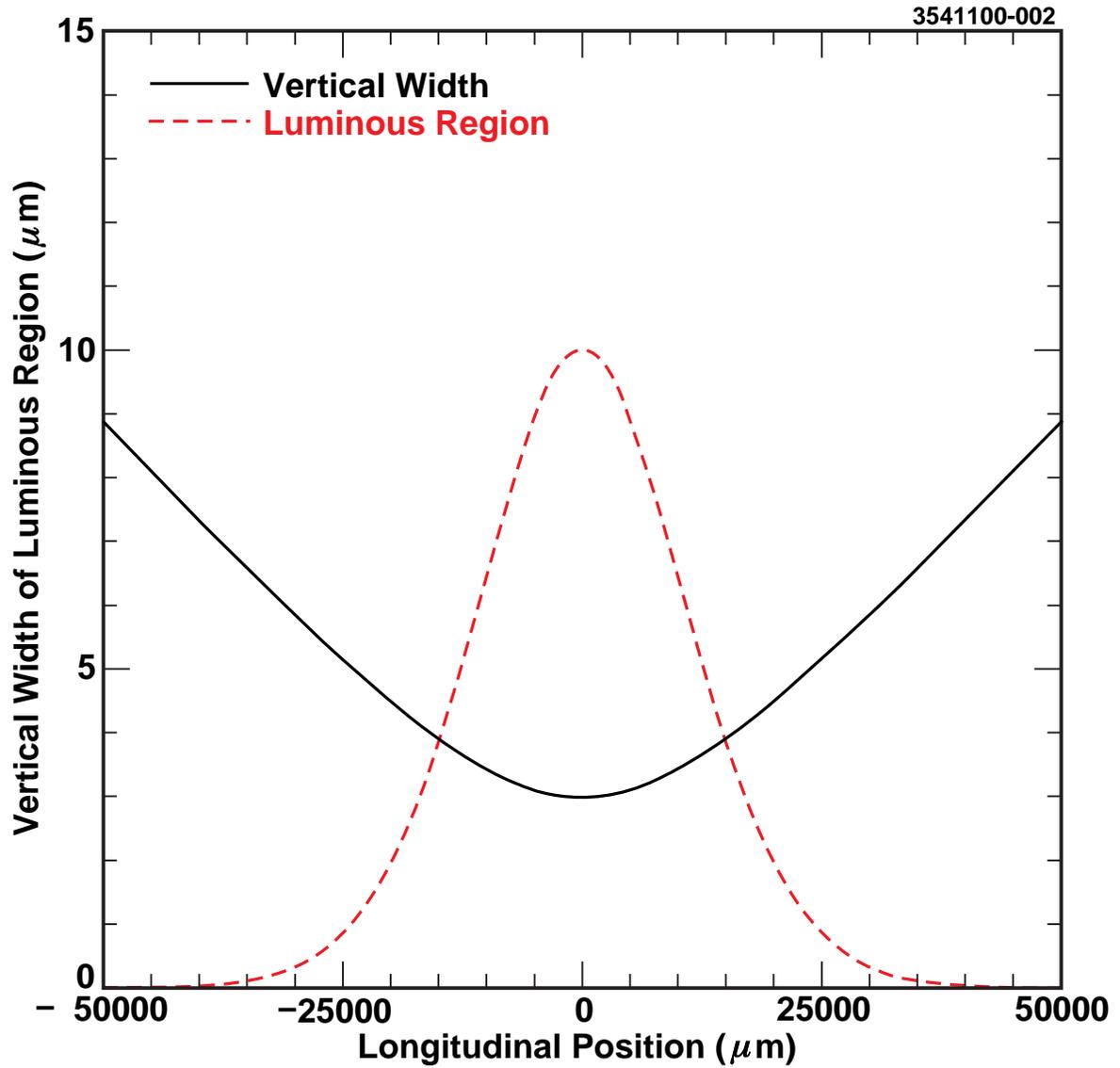,height=6.0in}
\caption{\label{fig:hourglass} Expected vertical width of the luminous 
region as a function of longitudinal position.  Also shown, in arbitrary 
units, is the expected longitudinal distribution of luminosity.}
\end{center}
\end{figure}

	Figure~\ref{fig:hourglass} shows $\sigma_y$ as given by
Equation~\ref{eq:houryres} using the beam parameters given in
Table~\ref{tab:beam}.  Also shown is the expected longitudinal
distribution of luminosity.
This is given by\cite{hourglass}
\begin{equation}
\frac{d{\cal L}}{dz} = {\cal L}_0 \frac{\exp{\left(\frac{-(z-z_{0 {\rm bunch}})^2}{\sigma_z^2}\right)}}
                   {(1 + \frac{(z-z_{0 {\rm beta}})^2}{\beta_x^{\ast 2}})^{1/2}
                    (1 + \frac{(z-z_{0 {\rm beta}})^2}{\beta_y^{\ast 2}})^{1/2}}\ ,
\label{eq:lumz}
\end{equation}
where $z_{0 {\rm bunch}}$ is the longitudinal position of the bunch-bunch
collision.  Note the longitudinal distribution of luminosity is
expected  to significantly depend on $\beta_y^\ast$, but the
$\beta_x^\ast$ dependence is negligible.  This is
due to the large size of $\beta_x^\ast$ as compared to $\sigma_z$.
Thus we expect a negligible hourglass effect 
in the horizontal size of the luminous region as a function of longitudinal 
position.  This is also why we consider only one value for $z_{0 {\rm beta}}$
which could, in principle, be different for the horizontal and vertical
beta functions.

	Our goal is to measure the beam parameters, $\beta_y^\ast$, 
$\epsilon_y$, and incidentally $\sigma_z$.  We do this with a
simultaneous fit to the 
measured vertical width of the luminous region versus longitudinal position, 
and the longitudinal distribution of luminosity.  The vertical 
width depends on $\epsilon_y$, the longitudinal distribution on $\sigma_z$ 
and they both depend on $\beta_y^\ast$.

	CESR has been described in detail elsewhere.~\cite{CESR}
All the data used in this measurement are taken at an $e^+e^-$ collision
energy of 10.58 GeV, and with bunch currents in the range of 2.5
to 7.0 mA over a four month period in late 1998 and early 1999.  The 
CLEO detector has also been described in detail elsewhere.\cite{CLEO} 
All of the data used in this measurement are taken in the CLEO~II.V
configuration which includes a silicon strip vertex detector which
is crucial to the measurement of the luminous region.
This consists of 
three layers of silicon wafers arrayed in an octagonal geometry around 
the interaction point.  The first measurement layer is at a radius of 
2.3 cm and the wafers are read out on both sides by strips which are 
perpendicular to each other.  The readout strips have a pitch of about 
100$\ \mu$m and with charge sharing the detector has an intrinsic per 
point resolution of better than 20$\ \mu$m in both the transverse plane
and the longitudinal direction

	To obtain a resolution of order 10$\ \mu$m on the luminous region, 
we selected $e^+e^- \rightarrow \mu^+\mu^-$ events. These are easily 
selected in CLEO as events with two and only two tracks each with
momentum near the beam 
energy and a small energy deposit in the electromagnetic calorimeter.  
We chose tracks with 20 or more hits in the main drift chamber, and at least two 
silicon vertex detector hits in the transverse and longitudinal views.
We require that the tracks have opposite charge and 
that those used for the measurement of the 
luminous region have at least three silicon vertex detector
hits in one of the two views.  

\begin{figure}
\begin{center}
\epsfig{file=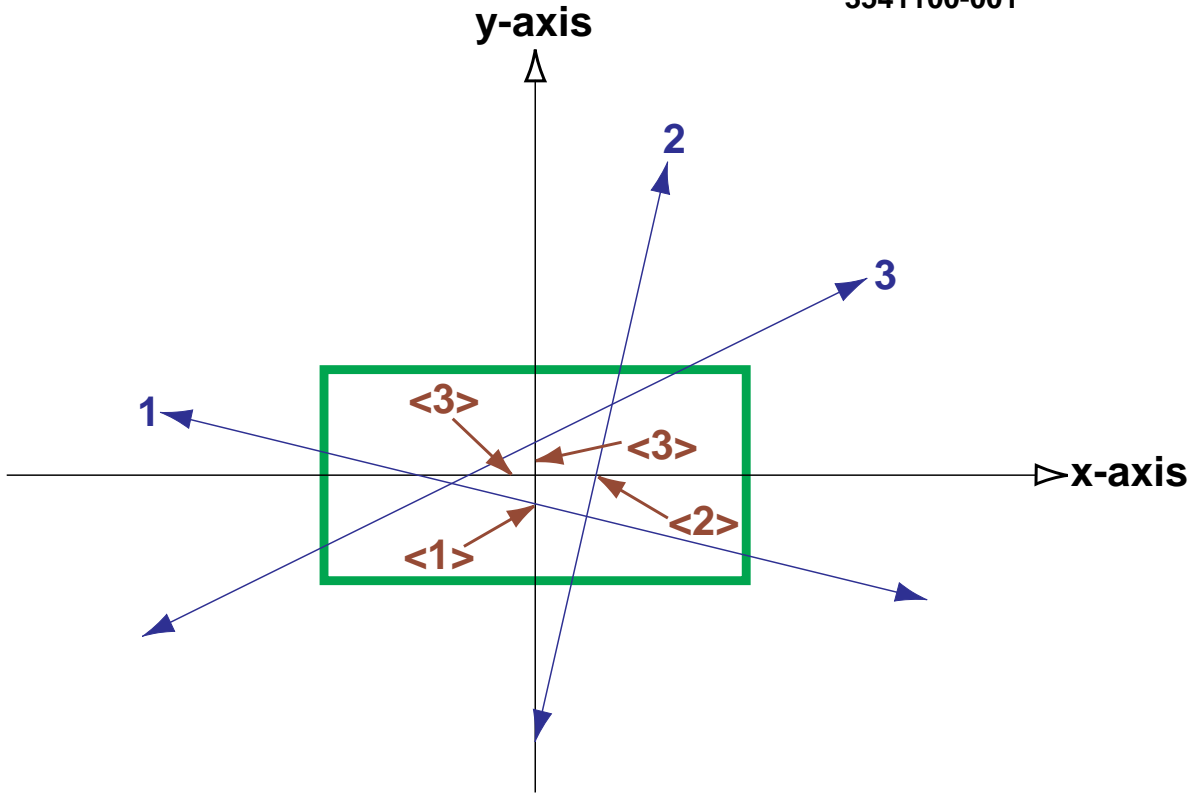,height=4.45in}
\caption{\label{fig:boxmethod}
An ensemble of stiff tracks passing through the box 
allows for a precision measurement of the luminous region.
For example, the track labeled~1 only gives a useful measure
of the vertical position of the luminous region, as indicated.
Track~1 crosses the the entire horizontal extent of the box and
its average horizontal position is simply the center of the box, 
rather than the center of the luminous region. Similarly track~2 
only measures the horizontal position, while track~3 measures both 
the horizontal and vertical positions of the luminous region.}
\end{center}
\end{figure}

	We implement a method, called the ``box technique,'' to obtain 
measurements of the beam parameters and the resolution.  
Figure~\ref{fig:boxmethod}
shows how this technique is implemented.  First, the size and location 
of the luminous region are obtained from run average data using hadronic 
events.\cite{dybeta}  A three-dimensional box is then centered about the 
measured center of the luminous region with sides ten times the measured 
widths of the luminous region.  The average position 
of a track passing through the box is found.
From an ensemble of such positions the size and shape 
of the luminous region is measured.

	Tracks that are parallel to an axis of interest are the
most useful for measuring the luminous region.  Tracks that are
perpendicular to an axis cross the full length of the box in
that direction and give no information about the luminous region.
We select appropriate tracks by cutting on the direction cosines.  
Essentially, these cuts are determined by the size of the luminous region, 
which is roughly 10 $\mu$m vertically, 300 $\mu$m horizontally, and 
10000 $\mu$m longitudinally.  Thus a tight cut of $|p_y/p| \equiv
|\cos\theta_y| < 0.1$  is needed to measure the vertical luminous region,
a looser cut of  $|\cos\theta_x| < 0.3$ for horizontal, and
$|\cos\theta_z| < 0.7$ for 
longitudinal.  For tracks with large $|\cos\theta_z|$, the resolution 
degrades, and the $|\cos\theta_z|$ cut of $0.7$ is also used on tracks 
to make vertical and horizontal measures.  Because of these direction 
cosine cuts, a single track can measure, at most, two dimensions.

	We tested this method using over 100,000 
$e^+e^- \rightarrow \mu^+\mu^-$ simulated events.  To measure the change
in the vertical size of the luminous region as expected
from the hourglass effect, a constant vertical 
resolution is necessary.  Thus we made selections in the simulated data 
to test the stability of the resolution and found significant dependences 
only on tracks with large values of $|\cos\theta_z|$, which are eliminated 
by the direction cosine cut discussed above, and on
the $|\cos\theta_y|$ of the tracks.  The $0.1$ 
cut on the $|\cos\theta_y|$ is a compromise between a 
smaller cut value with improved resolution, and a larger value with 
increased statistics.  Thus the resolution on the vertical luminous region 
is expected to be $26.4 \pm 0.4 \pm 1.5\ \mu$m with the first 
error due to the statistics of the simulation sample and the second due to 
the sharp dependence on the $|\cos\theta_y|$ cut.  This vertical resolution is 
small enough and there are sufficient data to obtain a statistically useful 
sample to measure the size of the luminous region at large longitudinal 
positions.  These can be compared with similar measures dominated by
the resolution taken at small longitudinal positions.  In the analysis,
we extract the resolution from the data itself, 
thus there is no dependence on the prediction from the simulation.

\begin{figure}
\begin{center}
\epsfig{file=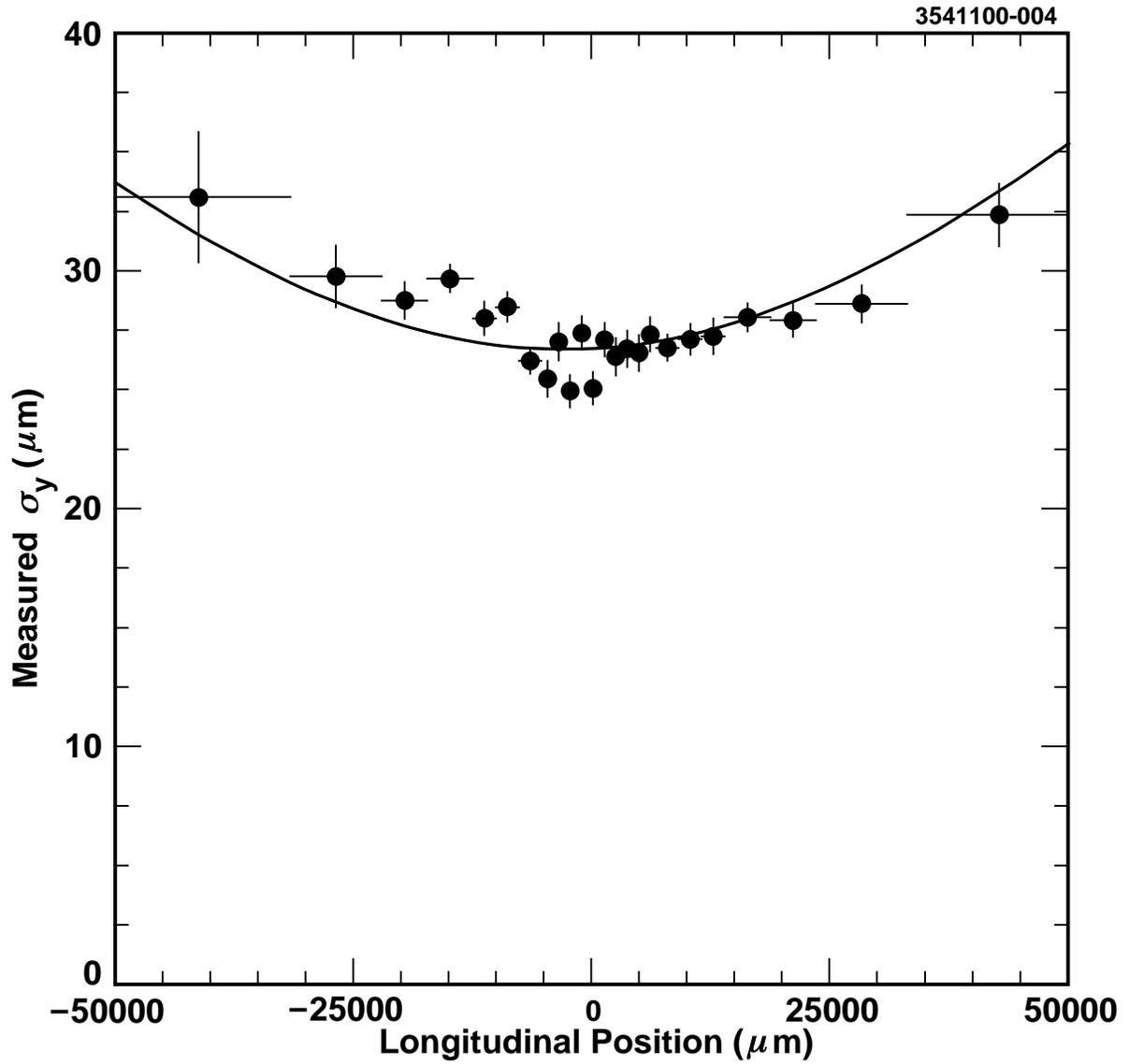,height=6.0in}
\caption{\label{fig:vert_dist} The vertical width of the luminous 
region as a function of longitudinal position.  The line shows the fit
discussed in the text.}
\end{center}
\end{figure}

	Figure~\ref{fig:vert_dist} shows the vertical width of the 
luminous region as a function of its longitudinal position. 
This plot clearly shows the vertical distribution growing away from the 
center.  This is evidence of the hourglass effect.
We have also repeated this procedure for the horizontal width.
We observe a horizontal width of 296 $\mu$m and see no
significant hourglass effect.  These observation both agree with
our expectation.

\begin{figure}
\begin{center}
\epsfig{file=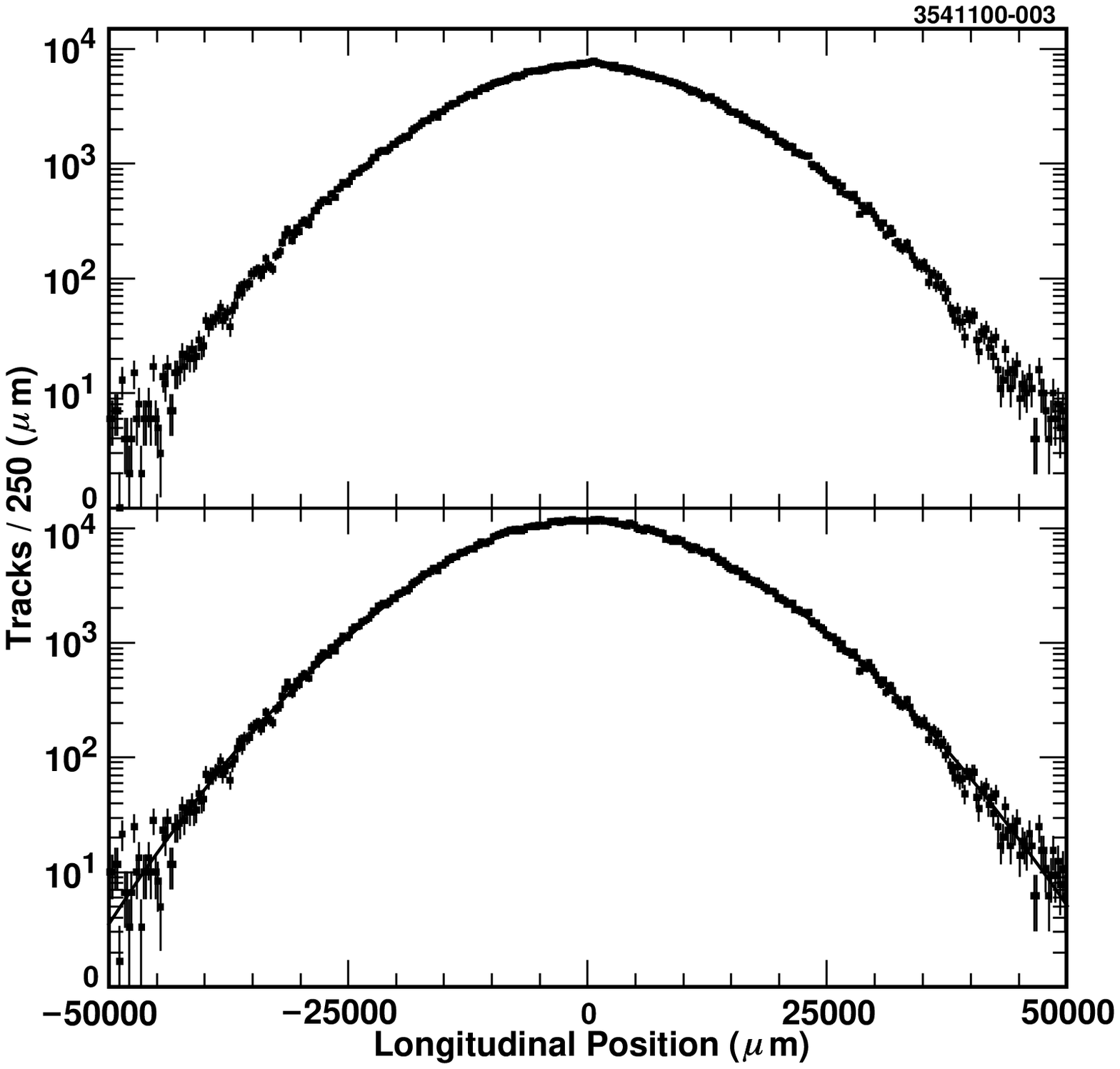,height=6.0in}
\caption{\label{fig:long_dist} The longitudinal distribution of luminosity.
The top plot shows the raw distribution.  Note
the slight enhancement near zero caused by detector geometry
as discussed in the text.  The bottom plot shows the efficiency
corrected distribution with the fit discussed in the text superimposed.}
\end{center}
\end{figure}

	When measuring the longitudinal distribution of luminosity
as a function of the longitudinal position, we see a sharp
enhancement in the distribution for small values of
the longitudinal position.  This enhancement is caused by 
the existence of a non-sensitive region in the center of the detector
which greatly diminishes the chance for tracks passing through this
region to be accepted for analysis.  This geometric effect
is accurately modeled in our simulation, and we use 
it to extract a longitudinal position dependent correction
for the longitudinal distribution of luminosity.
Applying this efficiency eliminates large systematic effects 
in our extraction of $\beta_y^\ast$.  The longitudinal 
distribution of luminosity both before and after
the efficiency correction is shown in Figure~\ref{fig:long_dist}.

	To extract the beam parameters, we fit Figure~\ref{fig:vert_dist} to 
Equation~\ref{eq:houryres}, including resolution smearing.  Unfortunately, 
such a fit does not give a useful measurement of any of the beam parameters, 
and has nearly 100\% correlations among $\beta_y^\ast$, $\epsilon_y$, and 
the resolution.  We take advantage of the dependence of the longitudinal 
distribution of luminosity on $\beta_y^\ast$ as given in 
Equation~\ref{eq:lumz}, and perform a simultaneous maximum likelihood fit 
to Figures~\ref{fig:vert_dist} and \ref{fig:long_dist} to
Equations~\ref{eq:houryres} and~\ref{eq:lumz}.
This simultaneous fit gives us an additional 
constraint on $\beta_y^\ast$ which breaks the correlations among the
parameters.  

	In this fit we fix the value of $\beta_x^\ast$ to
417500 $\mu$m based on our observation of the horizontal width
of the luminous region and the expected $\epsilon_x$ given
in Table~\ref{tab:beam}.  The results are not sensitive to
the exact value of $\beta_x^\ast$ used in the fit and change
negligibly if $\beta_x^\ast$ is left to float.  If $\beta_x^\ast$
is left to float the fit returns a value consistent with
417500 $\mu$m but with large errors of $\pm$500000 $\mu$m.

	The resolution of the box technique on the longitudinal position 
of the event production point is better than $40\ \mu$m.  This is negligible
in comparison with the longitudinal size of the luminous region, which is over 
one centimeter.  In fact, the box technique can be used to make a very high 
precision measurement of the bunch length, $\sigma_z$. As discussed in 
Reference \cite{bunchl} the bunch length in CESR is seen to depend on 
the bunch current and the bunches are asymmetric with their heads being 
narrower than their tails.  Due to the collision of the two
bunches this single bunch asymmetry is washed out in the longitudinal
distribution of the luminous region.  The luminosity distribution would only
be distorted away from a Gaussian shape if the single bunch asymmetry were
an order of magnitude larger than the observed $\sim 5$\%.  If we allow an
asymmetry in the luminosity
distribution, we observe none with a $\pm 1$\% accuracy.  This also confirms
our need for an efficiency correction which takes out a $1.5$\% asymmetry
in the raw data. 

	We have tested this simultaneous fit procedure with the simulation,
and  expect that our fit is able to measure the input beam parameters and the
resolution.  We also derived expectations on the errors and
correlations the fit should return based on our data statistics.

	Table~\ref{tab:hourfit} shows the results of the fit to the data.
\begin{table}
\caption{The results of
the simultaneous fit to the data distributions for
the vertical width of the luminous region as function
of longitudinal position and the longitudinal distribution.
Only statistical errors are shown.}
\begin{center}
\begin{tabular}{|c|c|} \hline
Parameter            & Fitted Value ($\mu$m) \\ \hline
$\beta^\ast_y$       & $15699 \pm 138$ \\
$\epsilon_y$         & $0.0060 + 0.0047 - 0.0042$ \\
$\sigma_z$           & $19288 \pm 38$ \\
resolution           & $25.8 \pm 1.7$ \\
$z_{0{\rm beta}}$      & $ -2885 \pm 101$ \\ 
$z_{0{\rm bunch}}$     & $ 852.8 \pm 34.7 $ \\ \hline 
\end{tabular}
\end{center}
\label{tab:hourfit}
\end{table}
From the fit we observe some large correlations.  These are  
between $\beta_y^\ast$ and $\sigma_z$, between $\epsilon_y$ and the
resolution, and between $z_{0 {\rm beta}}$ and $z_{0 {\rm bunch}}$
These correlations are 
-88\%, -75\%, and -85\% respectively.  They are of the 
size predicted by our tests on simulated data.  All other correlations are 
smaller than 40\% in magnitude..  

	These are in good agreement with 
the errors expected from the simulation study, the CESR 
beam parameters of Table~\ref{tab:beam}, and streak
camera observations.\cite{bunchl}  Note that 
we obtain very accurate measures of  $\beta_y^\ast$ and $\sigma_z$, along 
with a resolution from the data consistent with our expectation
of $26.4 \pm 1.6\ \mu$m from the simulation, but only a 1.4 standard 
deviation measure of $\epsilon_y$.
The results follow our expectations from the dynamic effects caused by
the non-zero bunch current.  The value for $\beta^\ast_y$ is
lower than the zero bunch current value, but not as small as the
lowest recorded.  The value for $\epsilon_y$ is not measured
well enough to make a meaningful test, but it is certainly consistent
with an increase.  The $\sigma_z$ is increased by 6.6\% which is
consistent with the streak camera observations.\cite{bunchl}
The difference between $z_{0{\rm beta}}$ and $z_{0{\rm bunch}}$,
$-3740 \pm 130\ \mu$m, is consistent with known strength differences
between the final focus quadrupoles and alignment tolerances
between final focus elements and RF cavities which respectively
determine the longitudinal position of the beta waist and the
center of the bunch collision.

	In an attempt to improve the measure of $\epsilon_y$, we 
repeat the data fit with the resolution fixed to $26.4\ \mu$m, as 
predicted by the simulation.  This fit does give a slightly improved
measurement of $\epsilon_y$ of $0.0049\pm0.0028\ \mu$m with the other 
parameters changing negligibly.  When we vary the fixed resolution by 
$\pm 1.6\ \mu$m as indicated by the simulation studies, this introduces 
an error of $\pm 0.0028\ \mu$m on $\epsilon_y$.  The combined error of 
$\pm 0.0040\ \mu$m on $\epsilon_y = 0.0049\ \mu$m is consistent with, 
but not a substantial improvement over the results of 
Table~\ref{tab:hourfit}.  We prefer to quote results for the fit where the 
resolution is left floating.

	We varied the standard fit to test its robustness.  We excluded
positive $z$, negative $z$, small $z$, and large $z$ data from the fit.
A $\chi^2$ fit is used rather than a 
likelihood fit.  The only parameters that show significant disagreement 
with the standard result are $\beta_y^\ast$ and $\sigma_z$.
Other facets of the analysis are varied and the procedure 
is repeated to estimate other systematic effects.  Cuts on the direction
cosines are varied, $\pm 0.01$ on $\cos\theta_y$ and $\pm 0.1$ on 
$\cos\theta_z$, we relax the three silicon vertex detector
hits in one view to a looser two hit per view requirement, we vary
the procedure for applying the efficiency for the luminosity
as a function on the longitudinal position, and use the 
simulation efficiency without errors as an estimate of the effects of 
our limited simulation statistics. 
Some of the measured vertical widths are not consistent with
their longitudinal neighbors as can be seen in Figure 3.  This indicates
a systematic error of about two microns in the extraction of the  
widths with the box method.  Including this error has a negligible
impact on the results of the fit, increasing the statistical   
errors by less than 10\%.  For all these variations, the change 
in the central values of the beam parameters from the standard procedure 
is taken as the systematic effect. 
The combined effects of these variations
result in a systematic error of 
$\pm 460\ \mu$m on $\beta_y^\ast$, $\pm 0.0019\ \mu$m on $\epsilon_y$  
and $\pm 200\ \mu$m on $\sigma_z$. 

\newpage

	In conclusion, we use a new box technique to measure the size of 
the luminous region of CESR at the CLEO interaction region.  This new 
method takes advantage of the hit resolution in the CLEO~II.V 
silicon vertex detector and the well understood CLEO charged particle
tracking system in $e^+e^- \to \mu^+\mu^-$ events to precisely measure
the size of the luminous region.  The technique has a 
resolution of $25.5 \pm 2.0\ \mu$m which we extract from a fit to the data.
The excellent 
resolution of the box technique, combined with the large size of the 
CLEO II.V data set, allows us to make a clear observation of the hourglass 
effect, the increase in the size of the
luminous region away from the focal 
point.  The technique leads to measurement of the CESR beam parameters:
\begin{eqnarray}
\beta_y^\ast &  = & (15700 \pm 140 \pm 460)\ \mu{\rm m}, \\
\epsilon_y   &  = & (0.0060 \pm 0.0045 \pm 0.0019)\ \mu{\rm m},\\
\sigma_z     &  = & (19290 \pm 40 \pm 200)\ \mu{\rm m}.
\end{eqnarray}
where the first error is statistical and the second is systematic,
in a simultaneous fit of the vertical width of the luminous region as a
function of the longitudinal position and of the longitudinal distribution
of the luminosity.  Note that these measurements
imply that the vertical width of the CESR beam at the
CLEO collision point is $6.9 \pm 2.8\ \mu$m. 
These measurement are in good agreement
with the expectations of the theoretical CESR lattice taking
into account the dynamic effects caused by the non-zero bunch
current.  This technique
provides a non-invasive way to measure beam parameters at
the collision point, but it requires the comparatively rare
$e^+e^- \to \mu^+\mu^-$ events and a well understood detector tracking
system.

\end{document}